\documentclass[aps,preprint,groupedaddress]{revtex4}
\usepackage{graphicx}
\usepackage{amssymb}
\usepackage{wasysym}
\usepackage{pifont}
\begin{document}

\title{Magnetic Reversal in Nanoscopic Ferromagnetic Rings} 
\author{Kirsten Martens}
\altaffiliation[Present address:]{Institut f\"ur Theoretische Physik, Universit\"at
Heidelberg, Philosophenweg 19, 69120 Heidelberg.}
\author{D.L.~Stein}
\altaffiliation[Present address:]{Department of Physics and Courant
  Institute of Mathematical Sciences, New York, NY, 10003.}
\affiliation{Department of Physics, University of Arizona, Tucson, Arizona 85721}
\author{A.D.~Kent}
\affiliation{Department of Physics, New York University, New York, NY 10003}


\begin{abstract}
We present a theory of magnetization reversal due to thermal fluctuations
in thin submicron-scale rings composed of soft magnetic materials.  The
magnetization in such geometries is more stable against reversal than that
in thin needles and other geometries, where sharp ends or edges can
initiate nucleation of a reversed state.  The $2D$ ring geometry also
allows us to evaluate the effects of nonlocal magnetostatic forces.  We
find a `phase transition', which should be experimentally observable,
between an Arrhenius and a non-Arrhenius activation regime as magnetic
field is varied in a ring of fixed size.
\end{abstract}

\pacs{PACS numbers: 05.40.-a, 02.50.Ey, 75.60.Jk}

\maketitle

\section{Introduction}
\label{sec:intro}

The dynamics of magnetization reversal in submicron-sized, single-domain
particles and thin films has attracted much attention given its importance
in information storage and other magnetoelectronic applications.  The
problem can be approached by stochastic methods: in the classical regime
(typically at temperatures above $\sim 1^\circ K$) the magnetization
dynamics is governed by the Landau-Lifschitz-Gilbert equation~\cite{LDE80}
perturbed by weak thermal noise.  The classical N\'eel-Brown
theory~\cite{Neel49,Brown63} of thermally induced reversal assumed a
spatially uniform magnetization and uniaxial anisotropy.  Experimental
confirmation of this theory has been provided for certain simple
single-domain systems (15-30~nm Ni, Co, and Dy
nanoparticles)~\cite{Wern97a}.

There nevertheless remain fundamental open questions, especially when there
is spatial variation of the magnetization density
~\cite{Braun93,BNR01,BB98,LZ03}. While in small particles that are
spherical or nearly so, as in~\cite{Wern97a}, the N\'eel-Brown theory
appears to work reasonably well, it appears to break down for elongated
particles, thin films, and other geometries, which exhibit far lower
coercivities than predicted~\cite{Wern96}.

Braun~\cite{Braun93} made an initial step by studying the effects of
spatial variation of magnetization density on magnetic reversal in an
infinitely long cylindrical magnet.  However, Aharoni~\cite{Aharoni96}
pointed out that the energy functional employed neglected important
nonlocal magnetostatic energy contributions, invalidating the
result. Further, for submicron-scale magnets with large aspect ratio,
finite system effects are likely to play an important role; for example,
simulations~\cite{BNR01,ERV03b} indicate that magnetization reversal in
cylindrical-shaped particles proceeds via propagation and coalescence of
magnetic `end caps', nucleated at the cylinder ends. Both of these issues
are addressed in~\cite{Braun99}.

Here we consider a geometry that avoids these difficulties: an effectively
two-dimensional annulus.  Such systems have recently received increasing
attention~\cite{Zhu00,Rothshort}.  They are typically constructed of soft
magnetic materials (quality factor $Q\sim O(10^{-2})$), such as Fe, fcc Co,
or permalloy, have radii of order $(10^2-10^3)$~nanometers and thicknesses
of order 10~nm or less.

Our interest in these systems is twofold.  The first is technological:
because the magnetic bending length is much smaller than the typical system
size, there are two oppositely polarized stable states, each with
magnetization vector pointing everywhere along the circumferential
direction; they are degenerate in the absence of an external magnetic
field. But a current running along the ${\bf{\hat z}}$-direction through
the center, with ${\bf{\hat z}}$ the direction normal to the annulus plane,
generates a circumferential magnetic field breaking the degeneracy.  By
switching the direction of the current, the relative stability of the two
states is switched.  (A slightly different method, but with similar wire
dimensions and current magnitudes, was used in~\cite{Zhu00}).  The utility
of such a system as an information storage device depends on the
magnetization state being relatively long-lived against thermal
fluctuations, even at relatively high temperatures.  Unlike the cylindrical
particle, the micromagnetic ring has no ends where nucleation is easily
initated, making its magnetic state more stable against thermally induced
reversal.

The second is physical: by developing a theory for thermally induced
reversal that can be solved analytically, we are able to extract a number
of interesting qualitative features that would be more difficult to uncover
numerically and which should apply also to more complicated situations.
While several important quantitative features require a numerical
treatment, we show below that our most important qualitative findings are
robust (see in particular the Discussion section).

One of our central predictions is that a type of phase transition occurs in
the thermally induced reversal rate, and more importantly, that it can be
realized experimentally.  The possibility of such a transition in classical
stochastic field theories (which we show below includes the physical
problem of interest here) was first noted in~\cite{MS01b}, as system size
was varied in a symmetric Ginzburg-Landau double-well $\phi^4$ potential.
It was further shown in~\cite{Stein03} to apply more generally to
asymmetric systems as well.  In the present case, the transition depends on
{\it two\/} parameters: the system size and the strength of the applied
magnetic field.  Although the former cannot be continuously varied, the
latter can, facilitating experimental tests of the predicted transition.
In particular, we will show that as magnetic field varies for a ring of
fixed size, there should be a transition from a regime where activation is
Arrhenius to one where it is non-Arrhenius.

A preliminary account of this work has appeared in~\cite{MSK05}.

\section{The Model}
\label{sec:model}

We consider an annulus of thickness $t$, inner radius $R_1$ and outer
radius $R_2$.  We confine our attention to rings satisfying $t\ll\Delta
R\ll R$, where $\Delta R = R_2-R_1$ and $R=(R_1+R_2)/2$.  A current run
through the center leads to an applied field ${\bf H_e}$ at $R$ in the
circumferential direction ${\bf\hat\theta}$; the small variation $\sim
O(\Delta R/R)$ of field strength with radius can be ignored.  As will be
discussed below, magnetostatic forces produce strong anisotropies, forcing
the magnetization vector to lie in the plane and preferentially oriented
parallel to the inner and outer circumferences.  We may therefore consider
magnetization configurations that vary only along the
${\bf\hat\theta}$-direction.

Suppose now that the system is initially in its metastable state; i.e.,
with magnetization vector ${\bf M}=-M_0\hat{\bf\theta}$.  We are interested
in determining the mean rate for thermal fluctuations to reverse the
magnetization to its stable direction.  We consider temperatures above
$1^\circ K$, where classical thermal activation can be expected to apply.
The magnetization dynamics are then governed by the Landau-Lifshitz-Gilbert
equation~\cite{LDE80}
\begin{equation}
\label{eq:LLG}
\partial_t{\bf M}=-\gamma[{\bf M}\times{\bf H}_{\rm
eff}]+(\alpha/{M_0})[{\bf M}\times\partial_t{\bf M}]\, ,
\end{equation}
where $M_0$ is the (fixed) magnitude of ${\bf M}$, $\alpha$ the damping
constant, and $\gamma>0$ the gyromagnetic ratio.  The effective field ${\bf
H}_{\rm eff}=-\delta E/\delta{\bf M}$ is the variational derivative of the
total energy $E$, which (with free space permeability $\mu_0=1$)
is~\cite{ERV03b,Aharoni00}:
\begin{eqnarray}
\label{eq:energy0}
E[{\bf M}({\bf x})]&=&\lambda^2\int_\Omega d^3x |\nabla{\bf
M}|^2+{1\over 2}\int_{{\bf R}^3}d^3x|\nabla U|^2\nonumber\\ &-&\int_\Omega d^3x {\bf
H}_{\rm e}\cdot {\bf M}\, ,
\end{eqnarray}
where $\Omega$ is the region occupied by the ferromagnet, $\lambda$ is the
exchange length, $|\nabla{\bf M}|^2=(\nabla M_x)^2+(\nabla M_y)^2+(\nabla
M_z)^2$, and $U$ (defined over all space) satisfies $\nabla\cdot(\nabla
U+{\bf M})=0$.  The first term on the RHS of Eq.~(\ref{eq:energy0}) is the
bending energy, the second the magnetostatic energy, and the last the
Zeeman energy.  Crystalline anisotropy terms are neglected, given their
negligibly small contribution; they can be easily included but will at most
result in a small modification of the much larger shape anisotropies, to be
discussed below.

%



\section{Energy Scaling and the Magnetostatic Term}
\label{sec:scaling}

The presence of the nonlocal magnetostatic term complicates analysis.
However, the quasi-$2D$ nature of the problem allows a significant
simplification, as shown by Kohn and Slastikov~\cite{KS03} in an asymptotic
scaling analysis that applies when the aspect ratio $k=t/R$ and the
normalized exchange length $l=\lambda/R$ are both small, and $l^2\sim
k\vert\log k\vert$.  These constraints restrict the range of ring
geometries to which our analysis applies.  Before discussing the KS result,
we recast the energy in dimensionless form, letting $X=x/R$ and similarly
for all other lengths, and let ${\bf h}={\bf H}_{\rm e}/(2M_0l^2)$.  Then,
integrating along the direction normal to the plane, we have for the
bending plus Zeeman energy contribution
\begin{equation}
\label{eq:energy1}
{E_b+E_z\over M_0^2R^3}=kl^2\int_\omega
d^2X\Big[(\nabla_{\scriptscriptstyle X}{\bf m})^2-2{\bf h}\cdot{\bf
m}\Big]\, ,
\end{equation}
where ${\bf m}={\bf M}/M_0$ is the normalized magnetization and $\omega$
represents the $2D$ surface with boundary $\partial\omega$.

Before analyzing these terms further, we examine the magnetostatic energy
contribution. The analysis of~\cite{KS03} showed that this asymptotically
separates into local bulk and surface terms:
\begin{eqnarray}
\label{eq:mag}
{E_{\rm mag}\over M_0^2R^3}&=&{1\over 2}k\int_\omega d^2X
m_z^2+(1/4\pi)k^2|\log k|\int_{\partial\omega} ds_X ({\bf m}\cdot {\bf\hat
r})^2\nonumber\\ &+&{1\over 2}k^2 \int_\omega d^2X |\nabla\cdot{\bf
m}|^2_{H^{-1/2}}\, ,
\end{eqnarray}
where $s_X$ is dimensionless arc length along the boundary, and the final
integral is the squared $H^{-1/2}$ Sobolev norm of $\nabla\cdot
m$~\cite{DeSimone02}.  With current technology, the orders of magnitude
$k\sim10^{-2}$ and $l^2\sim k|\log k|\sim 10^{-2}-10^{-1}$ are just
attainable.  Then the first term of Eq.~(\ref{eq:mag}) is larger than the
others by roughly two orders of magnitude, forcing $m_z=0$, and therefore
in this topology we can ignore fluctuations of ${\bf m}$ out of the plane
(we will discuss this further in Sec.~\ref{subsec:topology}).

The second term, like the first, is a (local) magnetostatic surface (or
shape anisotropy) term.  The third term represents a nonlocal magnetostatic
bulk energy. When nonzero, this term will be roughly an order of magnitude
smaller than the others (see Sec.~\ref{subsec:bulk}), and so to a first
approximation~\cite{notemixing} it can be dropped.  This will result, for
some values of ring size and external field, in an error of up to $10\%$
in the computation of the action, so we can hope at best for reasonably
good quantitative predictions of the logarithm of the escape rate.  As
noted in the Discussion section, however, the important qualitative
features uncovered by our analysis should remain unaffected.

\begin{figure}
\includegraphics[width=0.48\textwidth]{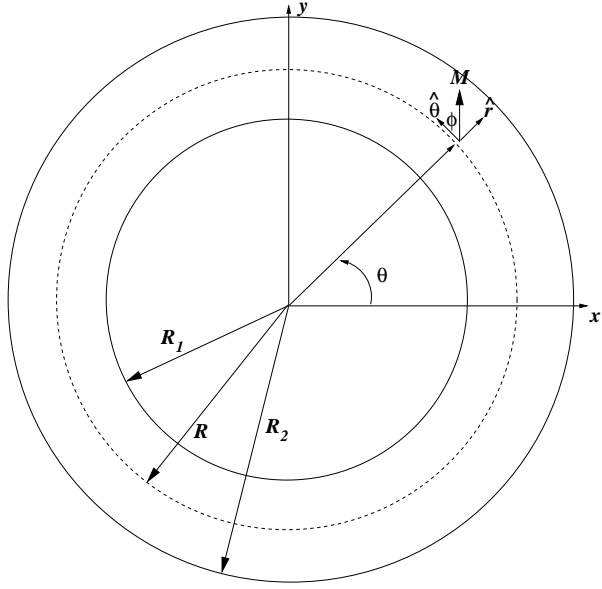}
\caption{Ferromagnetic annulus viewed from above, showing coordinates used
in text.}
\label{fig:annulus} 
\end{figure}

We use a locally varying coordinate system where the angle $\phi(s')$
measures the deviation of the local magnetization vector from the local
applied field direction; i.e., $\phi=0$ indicates that the local
magnetization is parallel to the local field, $\phi=\pi$ indicates that it
is antiparallel, and so on.  The parameter $s'=R\theta$ is the arc length
along the circumference.  The geometry and variables used are displayed in
Fig.~\ref{fig:annulus}.

The normalized magnetization vector can therefore be written, in
cylindrical coordinates, as ${\bf
m}=(m_r,m_\theta,m_z)=(\sin\phi,\cos\phi,0)$.  After integrating out the
radial coordinate the bending plus Zeeman energy becomes
\begin{eqnarray}
\label{eq:bending}
{E_b+E_z\over M_0^2R^3}=kl^2\Bigl[(\log{R_2\over R_1})\int_0^{2\pi}
d\theta\Big\{1+({\partial m_r\over\partial\theta})^2+({\partial
m_\theta\over\partial\theta})^2\nonumber\\+2(m_r{\partial
m_\theta\over\partial\theta}-m_\theta{\partial m_r\over\partial
\theta})\Big\}-{\Delta R\over R}\int_0^{2\pi}d\theta\ 2{\bf h}\cdot{\bf
m}\Bigr]\, .
\end{eqnarray}
The $m_r{\partial m_\theta\over\partial\theta}-m_\theta{\partial
m_r\over\partial \theta}$ term is a `winding number' of $\phi$ with respect
to the local direction; it gives zero in all configurations considered
here, but would give a nonzero contribution for uniform magnetizations,
e.g., ${\bf m}=\hat{\bf x}$.  For fixed $M_0$, it does not contribute to
the magnetization equations of motion given by Eq.~(\ref{eq:LLG}).

Finally, subtracting out constant terms and the first derivative term
(which gives zero contribution), noting that the boundary integral occurs
over both inner and outer radii, and rescaling lengths again gives
\begin{equation}
\label{eq:energy2}
{\cal E}=\int_0^{\ell/2}ds \Big[({\partial\phi\over\partial
s})^2+\sin^2\phi-2h\cos\phi\Big]\, ,
\end{equation}
where ${\cal E}=E/E_0=E/[2M_0^2R^2\Delta R\sqrt{c}kl^2]$,
$s=\theta\sqrt{c}$, $\ell=2\pi\sqrt{c}$, $h=\vert{\bf H_e}\vert/(2M_0l^2c)$,
and $c=(1/2\pi)(k|\log k|/l^2)(R/\Delta R)$. In deriving~(\ref{eq:energy2})
we used the fact that $\log(R_2/R_1)=\Delta R/R+O\Big[(\Delta R/R)^2\Big]$.
The error is negligible for the geometries considered here; for example,
with the ring parameters used in Fig.~\ref{fig:switching}, $\Delta R/R=1/5$
and $\log(R_2/R_1)=.20067$.  The parameter $c$ ($0<c<\infty$) depends on
the ring size and material properties; it represents the ratio of the
anisotropy energy scale to the bending energy scale, and determines the
width of a Bloch wall.

\subsection{Energetics and Topology}
\label{subsec:topology}

The scaling results of the previous section are useful insofar as they
provide results on how different contributions to the energy scale in the
thin-film limit.  Their effective application in a physical situation must
also take into account the geometry, and in our case, the topology of the
ferromagnetic particle under study.  Consideration of both of these aspects
provides a guide for considering what types of magnetization configurations
might be relevant in different thin-film geometries.

As one example, we consider the flat disk topology studied by
Shinjo~et~al.~\cite{SOHSO00} (corresponding to our geometry with $R_1=0$).
They studied magnetization configurations in permalloy disks of thickness
50 nm and diameters ranging from 0.3 to 1 micron.  Interestingly, they
observed vortex structures, particularly in the larger diameter samples.
The surface term energy in~Eq.~(\ref{eq:mag}) (which leads to the shape
anistotropy) is minimized by requiring the magnetization vector to remain
tangential to the surface (i.e., in the $\pm\hat\theta$-direction).  But
given the topology of the samples used here, this forces the interior
magnetization to do one of two things: either the magnetization {\it
magnitude\/} goes to zero at the center, or the magnitude stays mostly
constant but then the out-of-plane magnetization component $m_z\ne 0$ in
some interior region.  Either choice costs energy, but (when considered
over the same region) the first costs more than the second.

Given that there must be an out-of-plane magnetization component in the
disk topology studied by Shinjo et al.~\cite{SOHSO00}, the KS analysis can
determine the approximate lengthscale over which this occurs.  Comparing
the first term in Eq.~(\ref{eq:mag}) with the bending
term~(\ref{eq:bending}), we estimate that their respective energies are of
the same order when the the region where $m_z\ne 0$ is roughly of the order
of an exchange length, i.e., 10-20 nm.  This appears to be exactly what is
observed.  (Shinjo et al.~\cite{SOHSO00} don't provide an estimate for the
width of this region, noting only that they observe a contrast `spot' at
the center of each disk that corresponds to out-of-plane magnetization; see
Fig.~2 of their paper.)  Note that if the magnetization were to be
out-of-plane in a region much larger than this, Eq.~(\ref{eq:mag}) predicts
a prohibitively large energy cost.  (Simulation results consistent with
these conclusions appear in~\cite{UP93}.)

In our ring topology, however, an `outer vortex' configuration (i.e.,
magnetization circumferential at the outer boundary) does not require an
out-of-plane magnetization anywhere.  We can therefore ignore
configurations with $m_z\ne 0$, any of which are likely to have energies
larger than the configurations considered here.

\section{Transition in Activation Behavior}
\label{sec:transition}

The reversal rate $\Gamma$ due to thermal fluctuations at temperature $T$
is given by the Kramers formula $\Gamma\sim \Gamma_0\exp(-\Delta W/k_BT)$,
where the activation barrier $\Delta W\gg k_BT$ is simply the energy
difference between the (meta)stable and `saddle' states.  The latter is the
state of lowest energy with a single negative eigenvalue (and corresponding
unstable direction) of the linearized zero-noise dynamics.  Equivalently,
it is the configuration of highest energy along the system's optimal escape
path~\cite{HTB90}.  The rate prefactor $\Gamma_0$ is determined by
fluctuations about this optimal path, and its evaluation will be presented
in Sec.~\ref{subsec:prefactor}.

Stable, unstable, and saddle states are all time-independent solutions of
the LLG equations.  For fixed $M_0$, Eq.~(\ref{eq:LLG}) and the variational
equation ${\bf H}_{\rm eff}=-\delta E/\delta{\bf M}$ yield a nonlinear
differential equation that must be satisfied by any such time-independent
solution:
\begin{equation}
\label{eq:EL}
d^2\phi/ds^2= \sin\phi\cos\phi + h\sin\phi\, .
\end{equation}
There are three `constant' (i.e., $\phi$ is independent of $\theta$; these
remain nonuniform configurations because ${\bf m}$ varies with position)
solutions for $0\le h<1$: the stable state $\phi=0$ (${\bf m}=\hat\theta$);
the metastable state $\phi=\pi$ (${\bf m}=-\hat\theta$), and a pair of
degenerate unstable states $\phi=\cos^{-1}(-h)$, which constitute the
saddle for a range of $(\ell,h)$. The $\phi=0,\pi$ solutions are degenerate
when $h=0$, and the $\phi=\pi$ solution becomes unstable at $h=1$.  We
therefore confine ourselves to fields in the range $0\le h<1$.

We have also found a nonconstant (`instanton') solution~\cite{M04} of
Eq.~(\ref{eq:EL}), which we will see is the saddle for the remaining range
of $(\ell,h)$.  It is
\begin{equation}
\label{eq:instanton}
\phi(s,s_0,m)=2\cot^{-1}\Bigl[{\rm dn}\Big({s-s_0\over\delta}|m\Big){{\rm
sn}({\cal R}|m)\over{\rm cn}({\cal R}|m)}\Bigr]\, ,
\end{equation}
where dn$(\cdot|m)$, sn$(\cdot|m)$, and cn$(\cdot|m)$ are the Jacobi
elliptic functions with parameter $m$, $0\le m\le 1$~\cite{AS65}; $s_0$ is
an arbitrary constant arising from the rotational symmetry of the problem;
and ${\cal R}$ and $\delta$ are given by
\begin{eqnarray}
\label{eq:aux}
{\rm sn}^2({\cal R}|m)=1/m-h/2-(1/2m)\sqrt{m^2h^2+4(1-m)}&&\\
\delta^2={m^2\over 2-(m+\sqrt{m^2h^2+4(1-m)})}\, .&&
\end{eqnarray}
The period of the dn function equals $2{\bf K}(m)$, the complete elliptic
integral of the first kind~\cite{AS65}.  Accordingly, imposition of the
periodic boundary condition yields a relation between $\ell$ and $m$:
\begin{equation}
\label{eq:pbc}
\ell=2{\bf K}(m)\delta\, .
\end{equation}
In the limit $m\to 1$, corresponding to $\ell\to\infty$ at fixed $h$,
Eqs.~(\ref{eq:instanton})-(\ref{eq:pbc}) reduce to Braun's
solution~\cite{Braun93}.  In the limit $m\to 0$, ${\rm dn}(x|0)=1$, and the
instanton solution reduces to the constant state $\phi=\cos^{-1}(-h)$.  At
$m=0$, the critical length and field are related by
\begin{equation}
\label{eq:crit}
\ell_c=\pi\delta_c=2\pi/\sqrt{1-h_c^2}\, .
\end{equation}

\begin{figure}
\includegraphics[width=0.48\textwidth]{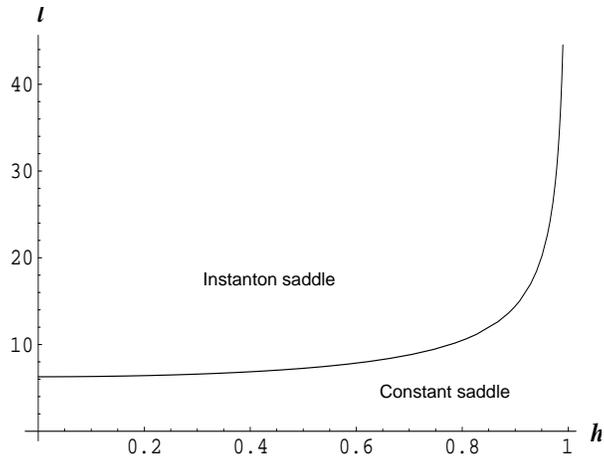}
\caption{The phase boundary between the two activation regimes in the
$(\ell,h)$-plane. In the shaded region the instanton state is the saddle
configuration; in the unshaded region, the constant state.}
\label{fig:boundary}
\end{figure}

The solution (\ref{eq:instanton})-(\ref{eq:pbc}) corresponds to a pair of
domain walls of width~$O(\delta)$. At fixed $h$, the constant configuration
is the saddle for $\ell<\ell_c$ and the instanton is the saddle for
$\ell>\ell_c$. This can be understood as follows: at fixed field, the
bending energy becomes sufficiently large at small $\ell$ so that the
constant state becomes energetically preferred. (There is a second
transition at even smaller $\ell$, where the bending energy becomes so
large that the magnetization lies along a single Euclidean direction
everywhere; we do not consider such small length scales here.)  Conversely,
at fixed $\ell$ the constant configuration is the saddle for $h>h_c$, and
the nonconstant for $h<h_c$.  (However, when $\ell\le2\pi$, the constant
configuration is the saddle for all $h$.) Here the Zeeman term dominates at
sufficiently large field, preferring a constant configuration, while at
smaller field the shape anisotropy energy dominates, preferring the
instanton configuration. The `phase boundary' (Eq.~(\ref{eq:crit})) is the
$m=0$ line in the $(\ell,h)$ plane, and is shown in
Fig.~\ref{fig:boundary}. We now compute the reversal rate in both regimes,
and examine how it is affected by the transition in the saddle state.

\section{The Reversal Rate}
\label{sec:reversal}

We turn now to a computation of the magnetization reversal rate $\Gamma$
due to thermal fluctuations at temperature $T$.  In equilibrium, it is
given, as noted in Sec.~\ref{sec:transition}, by the Kramers formula
$\Gamma\sim \Gamma_0\exp(-\Delta W/k_BT)$~\cite{HTB90}.  We first compute
the activation barrier $\Delta W$ for each saddle configuration.

\subsection{Activation Energy}
\label{subsec:activation}  

As noted earlier, the exponential dependence of the magnetization reversal
rate on temperature is given by $\Delta W$, the energy difference between
the saddle ($\phi_u$) and metastable ($\phi_s$) states (the notation arises
from the properties that the saddle is unstable along the longitudinal
escape direction, while the metastable state is locally stable in all
directions).  With the latter given by $\phi_s=\pi$, this is
\begin{eqnarray}
\label{eq:generg}
\Delta W/E_0 &=& {\cal E}[\phi_u]-{\cal E}[\phi_s=\pi]\\
&=&\int_0^{\ell/2}ds \left[({\partial\phi_u\over\partial s})^2
  +\sin^2(\phi_u)-4h\cos^2(\phi_u/2)\right]\nonumber\, .
\end{eqnarray}

\begin{figure}
\includegraphics[width=0.48\textwidth]{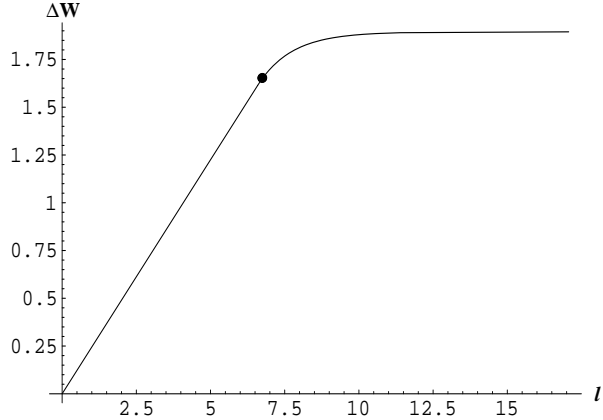}
\caption{Activation energy $\Delta W$ for fixed $h=0.3$ as $\ell$
varies. The dot indicates the transition from constant to instanton saddle
configuration.}
\label{fig:act1}
\end{figure}

When the constant state $\phi=\cos^{-1}(-h)$ is the saddle configuration,
it easily seen that $\Delta W = (1-h)^2\ell/2$. When the nonconstant, or
instanton, state is the saddle, the integral~(\ref{eq:generg}) must be
computed numerically.  However, it can be analytically computed in the
$m\to 0$ ($\ell\to\ell_c^+(h)$) limit, where one finds $\Delta W (m\to
0)\to (1-h)^2\ell/2$.  So the energy (and its first derivative, which can
also be computed) is continuous at $\ell_c(h)$.  Of course, the second
derivative is discontinous there.  Fig.~\ref{fig:act1} shows the activation
energy as a function of ring circumference for fixed field.

\begin{figure}
\includegraphics[width=0.48\textwidth]{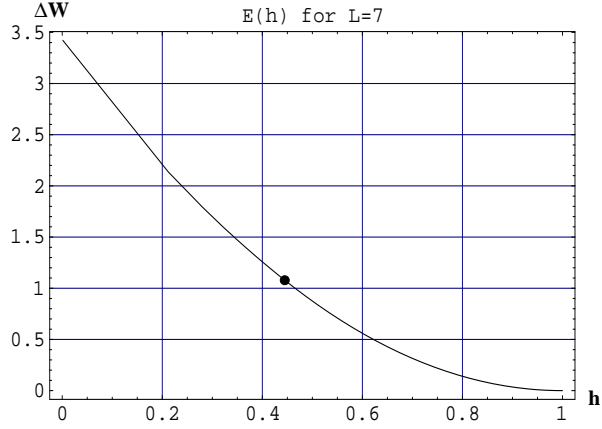}
\caption{Activation energy $\Delta W$ for fixed $\ell=7$ as $h$
varies. The dot indicates the transition from instanton to constant saddle
configuration.}
\label{fig:act2}
\end{figure}

The activation energy grows linearly with $\ell$ when the transition state
is constant; it becomes almost flat above $\ell_c$ at fixed $h$, because
the width of the domain walls remains essentially constant (cf.~Fig~4
of~\cite{Stein03}).  In Fig.~\ref{fig:act2} we show the activation energy
dependence on $h$ at fixed $\ell$, which is more relevant to experiment.

\subsection{Bulk Magnetostatic Energy Contribution}
\label{subsec:bulk}

We can now go back and check whether the contribution of the bulk
magnetostatic term is small compared to the bending energy.  This requires
an evaluation, or at least an estimate, of the $H^{-1/2}$ Sobolev norm of
the divergence of the reduced (i.e., in-plane) magnetization.  To do this,
we need to introduce some additional notation.  The $L^2(\omega)$ norm of a
quantity (say, the gradient of the reduced magnetization), is
\begin{equation}
\label{eq:L2}
\Vert\nabla{\bf m}\Vert_{{L^2}(\omega)}=\left[\int_\omega d^2x\ (\nabla{\bf m})^2\right]^{1/2}
\end{equation}
so that the dimensionless bending energy is simply $kl^2\Vert\nabla{\bf
m}\Vert^2_{L^2(\omega)}$.  From here on we shall abbreviate $L^2(\omega)$
to $L^2$ for ease of notation.

Formally, the $H^{-1/2}$ Sobolev norm of the magnetization divergence is
given by~\cite{DeSimone02}
\begin{equation}
\label{eq:Sobolev}
\Vert\nabla\cdot{\bf m}\Vert_{H^{-1/2}}=\Vert(\nabla)^{-1/2}(\nabla\cdot{\bf
m})\Vert_{L^2}\, .
\end{equation}

Its meaning becomes clearer through the use of Fourier transforms.  Define
the Fourier transform ${\hat f}({\bf\xi})$ of $f({\bf x})$ in the usual way:
$f({\bf x})=\int d^2\xi{\hat f}({\bf\xi})\ e^{i{\bf\xi}\cdot{\bf x}}$.  Then
\begin{eqnarray}
\label{eq:H12}
\Vert\nabla\cdot{\bf m}\Vert^2_{H^{-1/2}}&=&\int d^2\xi\ \left({\xi_1{\hat
m}_1+\xi_2{\hat m}_2\over\vert\xi\vert^{1/2}}\right)^2\nonumber\\
&=&\int d^2\xi\ {(\xi_1{\hat
m}_1+\xi_2{\hat m}_2)^2\over\vert\xi\vert}\, .
\end{eqnarray}

It now follows in a straightforward fashion that
\begin{eqnarray}
\label{eq:inequalities}
\Vert\nabla\cdot{\bf m}\Vert_{H^{-1/2}}&\le\Vert {\bf m}\Vert_{H^{1/2}}
&\le\Vert{\bf m}\Vert^{1/2}_{L^2}\ \Vert\nabla{\bf m}\Vert^{1/2}_{L^2}\nonumber\\\le&\Vert\nabla{\bf m}\Vert^{1/2}_{L^2}
\end{eqnarray}
where the last inequality follows because $\Vert\nabla{\bf
m}\Vert_{L^2}=1$.  Therefore, the bulk magnetostatic term is dominated by
the bending energy.

As noted earlier, the relevant scaling regime for the approach presented
here corresponds to $\ell^2\sim k\vert\log k\vert\sim 10^{-2}-10^{-1}$.
For the constant saddle configuration the maximal bulk magnetostatic energy
arises when $\phi_u=\pi/2$; this is also the maximum value of $\phi$ for
the saddle, corresponding to $h=0$.  For this configuration, and with ring
parameters used in Fig.~\ref{fig:switching}, an upper bound for the
magnetostatic bulk energy, computed using the
inequalities~(\ref{eq:inequalities}), is roughly an order of magnitude
smaller than the bending energy. As $h$ increases from 0, and
correspondingly $\phi_u\to\pi$, the magnetostatic bulk term decreases to
zero.

For the nonconstant, or instanton, saddle, the minimum value of $\ell$ is
$2\pi$.  At this lengthscale, an upper bound for the ratio of magnetostatic
bulk energy to bending energy varies roughly from 0.05 to 0.1 as $h$
varies; smaller numbers are found as lengthscale increases, justifying the
neglect of this term.

Qualitatively, the instanton configuration has nonzero divergence only over
a region of $O(\delta)$, which remains smaller than $O(1)$ except close to
$m=1$ ($\ell\to\infty$) and $h=1$.  The instanton configuration contributes
to the bending energy, however, over the entire ring.  It is therefore not
surprising that, in the appropriate scaling region, the instanton's
magnetostatic bulk energy is relatively small compared to its bending
energy.  This is in contrast to instanton configurations in the
cylinder~\cite{Braun93,Aharoni96}; there, while the region contributing to
a bulk divergence is $O(1)$, the same region supplies the entire
contribution to the bending energy as well, and so the magnetostatic
contribution cannot be neglected there.

\subsection{Rate Prefactor}
\label{subsec:prefactor}

The leading-order rate asymptotics are determined by the activation barrier
$\Delta W$; the {\it subdominant\/} asymptotics appear as the rate
prefactor $\Gamma_0$.  Because the magnitude of $\Gamma_0$ is controlled by
the extent of fluctuations about the optimal escape path, the prefactor is
considerably more difficult to calculate than $\Delta W$.  Although the
reversal rate is only linearly dependent on the prefactor, as opposed to
its exponential dependence on the activation barrier, the rate can still be
significantly affected by $\Gamma_0$, especially in the vicinity of a
transition in saddle configurations (cf.~Fig.~\ref{fig:prefactor}).
Moreover, an understanding of the prefactor is needed to study other
quantities of physical interest, such as exit location
distributions~\cite{MS97}.

The prefactor computation procedure is summarized in~\cite{MM89,HTB90} (see
also~\cite{Langer67,Schulman81}).  Consider a small perturbation $\eta$
about the metastable state, so that sufficiently close to it
$\phi=\phi_s+\eta$.  Then to leading order the time dependence of
fluctuations about the metastable state is given by
$\dot\eta=-{\bf\Lambda_s}\eta$, where ${\bf\Lambda_s}$ is the linearized
zero-noise dynamics at $\phi_s$.  Similarly ${\bf\Lambda_u}$ is the
linearized zero-noise dynamics around $\phi_u$. Then~\cite{MM89,HTB90}
\begin{equation}
\label{eq:prefactor}
\Gamma_0 = 
\frac1{2\pi}
\sqrt{\left|\frac{\det{\bf\Lambda_s}}{\det{\bf\Lambda_u}}\right|}
\,\,\left|\lambda_{u,0}\right|,
\end{equation}
where $\lambda_{u,0}$ is the single negative eigenvalue
of~${\bf\Lambda_u}$.  Its corresponding eigenvector is the direction along
which the optimal escape trajectory approaches the transition state. In
general, the determinants in the numerator and denominator of
Eq.~(\ref{eq:prefactor}) can separately diverge: they are typically
products of an infinite number of eigenvalues with magnitude greater than
one.  However, their {\it ratio\/}, which can be interpreted as the limit
of a product of individual eigenvalue quotients, is finite.

\subsubsection{Constant saddle}
\label{subsubsec:subcrit}

When $\ell<\ell_c(h)$, or equivalently $h>h_c(\ell)$, the saddle is the
pair of constant configurations $\phi=\cos^{-1}(-h)$, the prefactor can be
determined by direct computation of eigenvalues of the stable and unstable
states~\cite{Stein03}.  Linearizing around the stable state gives
\begin{equation}
\label{eq:phis}
\dot\eta=-{\bf\Lambda_s}\eta=-(-d^2/dx^2+1-h)\eta\, ,
\end{equation}
and similarly
\begin{equation}
\label{eq:phiu}
\dot\eta=-{\bf\Lambda_u}\eta=-(-d^2/dx^2+h^2-1)\eta\, 
\end{equation}
about the transition state.  The spectrum of eigenvalues corresponding
to ${\bf\Lambda_s}$ is
\begin{equation}
\label{eq:stablespectrum}
\lambda_n^s={4\pi^2n^2\over\ell^2}+1-h\qquad\qquad\qquad
n=0,\pm 1,\pm 2\ldots
\end{equation}
and the eigenvalues corresponding to ${\bf\Lambda_u}$ are
\begin{equation}
\label{eq:unstablespectrum}
\lambda_n^u={4\pi^2n^2\over\ell^2}+h^2-1\qquad\qquad\qquad n=0,\pm 1,\pm 2\ldots\, .
\end{equation}

This simple linear stability analysis justifies the claims that $\phi_s$ is
a stable state and $\phi_u$ a saddle.  Over the interval $[0,\ell_c)$ all
eigenvalues of ${\bf\Lambda_s}$ are positive, while all but one of
${\bf\Lambda_u}$ are.  Its single negative eigenvalue
$\lambda_0^u=-(1-h^2)$ depends on $h$ but is independent of $\ell$.

Putting everything together, we find
\begin{equation}
\label{eq:g0-}
\Gamma_0^-=\tau_0^{-1}{(1-h^2)\over\pi}{\sinh(\sqrt{1-h}\ell/2)\over\sin(\sqrt{1-h^2}\ell/2)}\,
,
\end{equation}
where $\tau_0^{-1}=\alpha\gamma E_0/M_0{\cal V}(1+\alpha^2)$, with ${\cal
V}$ the ring volume.  The rate includes a factor of 2 because the system
can escape over either of the saddles, which are rotationally equivalent
with respect to $\phi_s=\pi$.  

\begin{figure}
\includegraphics[width=0.48\textwidth]{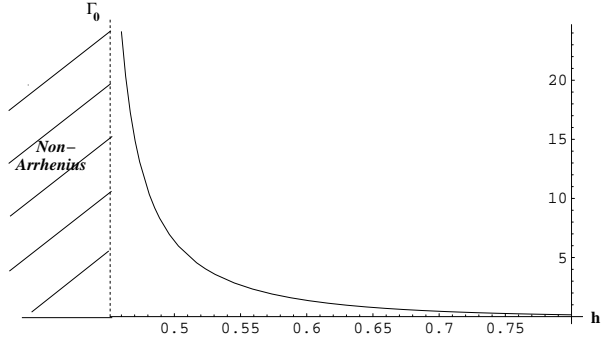}
\caption{The prefactor $\Gamma_0$ (in units of $\tau_0^{-1}$ vs.~$h$ when
$\ell=7$ on the `constant saddle' side of the transition.  The prefactor on
the `instanton saddle' side of the transition acquires an additional
temperature dependence, as discussed in the text.}
\label{fig:prefactor}
\end{figure}

The prefactor $\Gamma_0^-$ diverges at $\ell_c(h)$, or conversely
$h_c(\ell)$, as expected (cf.~Fig.~\ref{fig:prefactor}); in this limit,
$\Gamma_0^-\sim {\rm const}\times (\ell_c-\ell)^{-1}$ as $\ell\to \ell_c^-$
at fixed $h$, or as $(h-h_c)^{-1}$ as $h\to h_c^+$ at fixed $\ell$.  The
prefactor in this region for fixed $\ell$ as $h$ varies is plotted in
Fig.~\ref{fig:prefactor}.  The divergence arises from the vanishing of the
eigenvalue of a pair of degenerate eigenfunctions at the critical point.
This indicates the appearance of a pair of soft modes, resulting in a
transverse instability of the optimal escape trajectory as it approaches
the saddle.  The meaning and interpretation of the divergence is discussed
in detail in~\cite{Stein03}.  Near (but not at) the critical point the
prefactor formulae hold, but in a vanishing range of $T$ as $\ell_c$ is
approached.  Exactly at $\ell_c$ the prefactor is finite but non-Arrhenius
(with a different exponent than that in Eq.~(\ref{eq:pref2})). Inclusion of
higher-order fluctuations~\cite{Reznikoff04} about the saddle can be used
to compute the prefactor at criticality, and will be addressed elsewhere.
We return to the prefactor divergence in Sec.~\ref{sec:discussion}.

The independence of $\Gamma_0$ with respect to temperature leads to the
well-known exponential temperature dependence of the overall reversal rate.
By analogy with chemical kinetics, this exponential falloff of the rate is
often called `Arrhenius behavior'.

\subsubsection{Nonconstant (instanton) saddle}
\label{subsubsec:supercrit}

Computation of the determinant quotient in Eq.~(\ref{eq:prefactor}) is less
straightforward when the transition state is nonconstant, i.e., when
$\ell>\ell_c(h)$ or equivalently $h<h_c(\ell)$ .  An additional complication
follows from the translational degeneracy (energy invariance with respect
to choice of $s_0$) of the nonconstant state.  This implies a soft
collective mode in the linearized dynamical operator ${\bf\Lambda_u}$ of
Eq.~(\ref{eq:prefactor}), resulting in a zero eigenvalue for all
$h<h_c(\ell)$ (not to be confused with the vanishing of the lowest stable
eigenvalue of the saddles exactly at $h_c(\ell)$).  

To proceed, we use the McKane-Tarlie regularization
procedure~\cite{McKane95}, which allows the evaluation of
$\det'{\bf\Lambda_u}$, the functional determinant of the operator
$\bf{\bf\Lambda_u}$ with the zero eigenvalue removed.  We refer the reader
to~\cite{McKane95} for details, but sketch the main features here.

With periodic boundary conditions, it is formally the case that
\begin{equation}
\label{eq:regdet}
{\det'{\bf\Lambda_u}\over\langle \eta_1|\eta_1\rangle}
={\eta_2(s+\ell,s_0;m)-\eta_2(s,s_0;m)\over \eta_1(s,s_0;m)\det{\bf
H}(s,s_0;m)}
\end{equation}
where $\eta_1(s,s_0;m)$ and $\eta_2(s,s_0;m)$ are two linearly independent
solutions of ${\bf\Lambda_u}\eta_i=0$, $i=1,2$, $\langle
\eta_1|\eta_1\rangle=\int_{-\ell/2}^{\ell/2}ds\ y_1^2(s,0;m)$ is the square
of the norm of the zero mode and $\det{\bf H}(s,s_0;m)=\dot
\eta_2(s,s_0;m)\eta_1(s,s_0;m)-\dot \eta_1(s,s_0;m)\eta_2(s,s_0;m)$ is the
Wronskian; here a dot denotes a derivative with respect to $s$.  The
expression (\ref{eq:regdet}) is meaningful only as part of a determinant
quotient, as noted above.

The functions $\eta_1$ and $\eta_2$ can be found by differentiating the
instanton solution~(\ref{eq:instanton}) with respect to $s_0$ and $m$,
respectively; i.e., $\eta_1(s,s_0;m)=\partial\phi(s,s_0;m)/\partial s_0$
and $\eta_2(s,s_0;m)=\partial\phi(s,s_0;m)/\partial m$.  This yields
\begin{eqnarray}
\label{eq:eta1} 
\eta_1(s,s_0;m)\,\,=\,\,-{2m\over\delta}{\rm sn}({\cal R}\vert m) {\rm
cn}({\cal R}\vert m)\nonumber\\
\times{{\rm sn}({s-s_0\over\delta}\vert m){\rm cn}({s-s_0\over\delta}\vert m)\over
{\rm cn}^2({\cal R}\vert m)+{\rm sn}^2({\cal R}\vert m){\rm
  dn}^2({s-s_0\over\delta}\vert m)}
\end{eqnarray}
and
\begin{eqnarray}
\label{eq:eta2} 
\eta_2(s,s_0;m)=\,\,-{2\over{\rm cn}^2({\cal R}\vert m)+{\rm sn}^2({\cal
  R}\vert m){\rm dn}^2({s-s_0\over\delta}\vert m)}\times\nonumber\\
  \Bigl\{{m(s-s_0)\over\delta^2}{d\delta\over dm}{\rm sn}({\cal R}\vert m)
  {\rm cn}({\cal R}\vert m){\rm sn}({s-s_0\over\delta}\vert m){\rm
  cn}({s-s_0\over\delta}\vert m)\nonumber\\+{{\rm sn}({\cal R}\vert m) {\rm cn}({\cal
  R}\vert m)\over 2(1-m)}
\Big[{\rm sn}({s-s_0\over\delta}\vert m){\rm
  cn}({s-s_0\over\delta}\vert m)E({s-s_0\over\delta}\vert
  m)\nonumber\\-{\rm sn}({s-s_0\over\delta}\vert m){\rm
  cn}({s-s_0\over\delta}\vert m)(1-m)({s-s_0\over\delta})-{\rm
  sn}^2({s-s_0\over\delta}\vert m){\rm dn}({s-s_0\over\delta}\vert
  m)\Big]\nonumber\\+{\rm dn}({\cal
  R}\vert m){d{\cal R}\over dm}{\rm dn}({s-s_0\over\delta}\vert m)\nonumber\\\Bigr\}
\end{eqnarray}
where ${\bf E}(\cdot|m)$ is the incomplete elliptic integral of the second
kind~\cite{AS65}.

Inserting these solutions into Eq.~(\ref{eq:regdet}) yields
\begin{equation}
\label{eq:unstable}
\left|{\det'{\bf\Lambda_u}\over\langle
\eta_1|\eta_1\rangle}\right|={\delta^3
\left[(2m/\delta)(d\delta/dm){\bf K}(m) - {\bf K}(m) +
 {\bf E}(m)/(1-m)\right]\over 4m^2{\rm sn}({\cal R}\vert m) {\rm cn}({\cal
  R}\vert m){\rm dn}({\cal R}\vert m)d{\cal R}/dm} \, .
\end{equation}

Using a similar procedure, we find the corresponding numerator for the
determinant ratio in~(\ref{eq:prefactor}) to be
\begin{equation}
\label{eq:stable}
\det{\bf\Lambda_s}=4\sinh^2\left(\delta\sqrt{1-h}{\bf K}(m)\right)\, ,
\end{equation}
consistent with the numerator of Eq.~(\ref{eq:g0-}), obtained through
direct computation of the eigenvalue spectrum.  (Recall, though, that it is
only the {\it ratio} of the determinants that is sensible.)  This becomes
clearer by noting that the expressions in Eqs.~(\ref{eq:unstable}) and
(\ref{eq:stable}) are well-behaved for all finite $\ell>\ell_c$ ($m>0$).
While both expressions separately diverge as $m\to 1$, it is easily checked
that the divergences cancel.

As already noted, the rotational symmetry of the instanton state
(corresponding to the arbitrariness of the constant $s_0$) corresponds to a
`soft mode', resulting in appearance of a zero eigenvalue $\lambda_{u,1}=0$
of the operator ${\bf\Lambda_u}$.  The corresponding eigenfunction is
clearly $\eta_1$ given by~(\ref{eq:eta1}).  The appearance of a zero mode
corresponds to the zero rotational energy of the instanton solution: the
center of the domain wall pair can appear anywhere in the ring. This is in
contrast to the situation in the finite cylinder~\cite{M04}, where the
instanton is `pinned'.  The general procedure for including the correction
required due to removal of the zero eigenvalue is described by Schulman
\cite{Schulman81}.  The correction in our problem is an additional factor
of $2\ell\sqrt{\langle \eta_1|\eta_1\rangle/\pi k_BT}$, which vanishes as
$m\to 0$ (and thereby removes the divergence of the prefactor as $m\to
0^+$).

\begin{figure}
\includegraphics[width=0.48\textwidth]{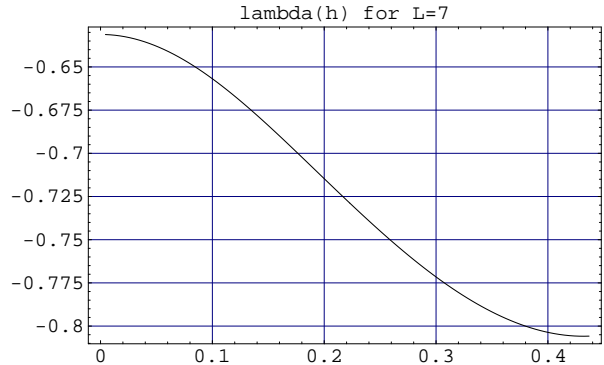}
\caption{Lowest eigenvalue $\lambda_{u,0}$ as a function of $h$ for $\ell=7$.}
\label{fig:eigenvalue}
\end{figure}

Finally, we need to compute the eigenvalue $\lambda_{u,0}$ corresponding to
the unstable direction.  With the substitution
$z=(s-s_0)/\delta$, the eigenvalue equation
${\bf\Lambda_u}\eta=\lambda\eta$ becomes
\begin{eqnarray}
\label{eq:eigenvalue}
0={d^2\eta\over dz^2}+\delta^2\eta\,\,\times&&\nonumber\\
\left\{ {(\lambda+1-h){\rm cn}^4({\cal R}\vert
  m)+2(\lambda+3){\rm sn}^2({\cal R}\vert m){\rm cn}^2({\cal R}\vert m)
  {\rm dn}^2(z\vert m) +(\lambda-1-h){\rm sn}^4({\cal R}\vert m){\rm
    dn}^4(z\vert m)\over({\rm cn}^2({\cal R}\vert m)+{\rm sn}^2({\cal R}\vert m){\rm dn}^2(z\vert m))^2}\right\}&&\, .
\end{eqnarray}
The lowest eigenvalue corresponds to a nodeless solution for $\eta$.  By
continuity it must tend towards $-1+h^2$ as $m\to 0^+$.  We have solved
Eq.~(\ref{eq:eigenvalue}) numerically for $\ell=7$; the result appears in
Fig.~\ref{fig:eigenvalue}.  The weak dependence on $h$ (and also $\ell$) is
typical.

Finally, we put all of the above results together to find the formula for
the prefactor {\it per unit length\/}:
\begin{eqnarray}
\label{eq:pref2}
\tau_0\Gamma_0^+/\ell=\vert\lambda_0(\ell,h)\vert
m(k_BT)^{-1/2}\sinh\Big[\delta\sqrt{1-h}{\bf
K}(m)\Big]\nonumber\\\times\Big[2m{\bf K}(m){d\log\delta\over dm}+{1\over
1-m}[{\bf E}(m)-(1-m){\bf K}(m)]\Big]^{-1/2}
\end{eqnarray}
where ${\bf E}(m)$ is the complete elliptic function of the second
kind~\cite{AS65}.  As noted above, $\lambda_0(\ell,h)$ is weakly dependent
on $h$ and $\ell$, and is $O(1)$ everywhere.

The most important qualitative feature to be noted from
Eq.~(\ref{eq:pref2}) is that the zero eigenvalue arising from the uniform
translation mode leads to non-Arrhenius behavior --- i.e., a $T$-dependent
prefactor --- everywhere on the low-field side of the transition.

Finally, we note that the eigenfunction $\eta_1$ given by~(\ref{eq:eta1})
has a single pair of nodes.  Because nodes arise in pairs, there must then
be only a single (nodeless) solution of lower eigenvalue than $\eta_1$.
But $\eta_1$ has zero eigenvalue, proving that the
solution~(\ref{eq:instanton}) has a single unstable eigenmode, and is
therefore a proper saddle.

\begin{figure}
\includegraphics[width=0.48\textwidth]{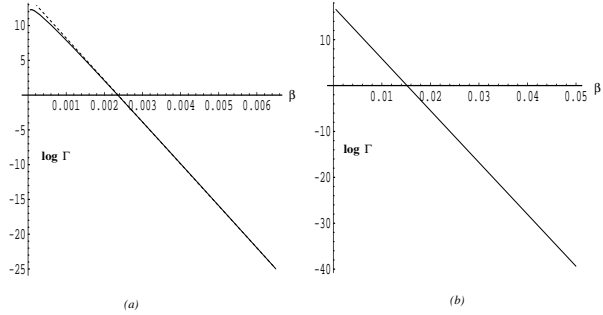}
\caption{Total switching rate (in units of $s^{-1}$) vs.~$\beta=1/k_BT$ (in
units of $^\circ K^{-1}$), at fields of (a) 60 mT (instanton saddle) and
(b) 72 mT (constant saddle). Parameters used are $k=.01$, $l=.1$,
$R=200$ nm, $R_1=180$ nm, $R_2=220$ nm, $M_0=8\times 10^5$ A/m (permalloy),
$\alpha=.01$, and $\gamma=1.7\times 10^{11}T^{-1}s^{-1}$. Deviation of
low-field switching rate in (a) from dashed line signals non-Arrhenius
behavior.}
\label{fig:switching}
\end{figure}

The above results allow one to find the overall reversal rate in any part
of the $(\ell,h)$ phase plane.  Results for a permalloy ring with given
dimensions are shown in Fig.~\ref{fig:switching}.

Among commonly used soft ferromagnetic materials, permalloy has the largest
magnetic exchange length.  The discussion of scaling in
Sec.~\ref{sec:scaling} suggests that the effects of nonlocal magnetostatic
terms are minimized with larger exchange lengths.  Where else might one
find magnetic materials with large exchange lengths?  Such materials would
require both low magnetization density and large exchange constants.  This
combination occurs naturally in certain ferrimagnets.  One example is
MgOFe$_2$O$_3$, which has an exchange length a factor of five larger than
that of permalloy. There are many examples of such materials that have been
prepared as polycrystalline thin films, and thus are soft magnets (i.e.,
have no or very weak magnetocrystalline anisotropies).  Such materials
might prove useful for experimental studies of the phenomena described in
this paper.

\section{Discussion}
\label{sec:discussion}

A theory of magnetization reversal in thin micromagnetic rings has been
presented.  Such systems are distinguished by their lack of edges or
corners where nucleation is easily initiated, leading to greater stability
of magnetization configurations and facilitating comparison of theory to
experiment.

By utilizing a scaling analysis~\cite{KS03} that uncovers a separation of
energy scales in the thin film limit, we are able to retain leading-order
terms that allow for an analytic solution of the relevant magnetization
configurations in the thermally-induced reversal problem.  The discarded
terms, in particular those corresponding to nonlocal magnetostatic energy
contributions, are shown to contribute no more than $O(10\%)$ to the energy
over most values in the $(\ell,h)$ phase plane.  A complete solution that
takes into account all terms must be numerical, and is planned for future
work.

Nevertheless, an analytic solution is highly useful and can uncover
information that may be difficult to extract from a numerical one. In
particular, we predict an unusual transition from Arrhenius to
non-Arrhenius activation behavior (cf.~Fig.~\ref{fig:switching}).  Our
analysis suggests that such a transition should now be observable
experimentally, by varying the externally applied magnetic field for rings
of fixed size.  A clear signature of such a transition would be the
observation of a crossover from Arrhenius to non-Arrhenius behavior as
field varies, as seen in Fig.~\ref{fig:switching}.  Because this requires
measurement of the prefactor, such an observation would require numerous
runs where reversal occurs.

Arrhenius behavior of magnetic reversal has already been found in several
systems and geometries.  In~\cite{Wern97a}, measurements of switching field
and waiting times on nearly spherical Ni, Co, and Dy nanoparticles found an
activation volume close to the particle volume, indicating a uniform
magnetization reversal (analogous to the constant saddle case here) and
confirming the N\'eel-Brown theory for these systems.  In contrast,
measurements on Ni wires with diameters 40-100 nm revealed an activation
volume considerably smaller than the particle volume, indicating a
nonuniform transition state (analogous to our instanton saddle).  Here,
too, Arrhenius switching behavior was found.  But wouldn't the arguments
given above imply that one should see non-Arrhenius behavior for these
wires?  No, because here the (roughly) cylindrical geometry with end caps
(and consequently the relevant boundary conditions on the magnetization),
lead to the absence of a uniform translation or rotational symmetry
present, and therefore no zero mode leading to a temperature-dependent
prefactor.  In fact, these observations support the robustness of our
conclusions for a wider variety of cases than considered in this paper.  We
will discuss this further below.

Before doing that, however, we wish to suggest a second experimental test
that may be easier to conduct: this is to measure the dependence of the
activation energy on mean radius $R$ for rings of identical composition.
On the Arrhenius side of the transition, where magnetization reversal
proceeds via a uniform rotation of the magnetization, the activation
barrier scales linearly with the ring size.  However, on the non-Arrhenius
side, where the instanton state governs the reversal, the activation
barrier is almost independent of ring size (see Fig.~\ref{fig:act1}).  In
this set of measurements, one may need to alter the applied field as ring
size varies to keep the system on one or the other side of the transition,
given that the critical field depends on $R$ (cf.~Eq.~(\ref{eq:crit})).

How robust are our predictions of a transition in activation behavior, and
in particular, can the neglected energy contributions wash out or obscure
the transition?  It is indeed possible, perhaps likely, that the details of
the transition close to the critical field (or circumference if field is
fixed) are sensitive to these terms.  In particular, the second-order
nature of the transition, and the corresponding divergence of the prefactor
(cf.~Fig.~\ref{fig:prefactor}), could disappear.  Inclusion of the
magnetostatic terms could even in principle change the transition from
second- to first-order, with a jump replacing the divergence in the
prefactor. Such first-order transitions have been predicted to occur in
thermally-induced conductance jumps in monovalent metallic
nanowires~\cite{BSS05}.

However, our central prediction, a transition from Arrhenius to
non-Arrhenius activation behavior, should be robust because it is due to
something much more fundamental: a rotationally invariant transition state
(our `constant' state $\phi=\cos^{-1}(-h)$) at high fields and a
rotationally non-invariant state (our instanton state~(\ref{eq:instanton}))
at low fields, with the crossover determined primarily through a
competition between the shape anisotropy arising from magnetostatic forces
and the Zeeman energy arising from the external field.  In fact, the
discussion in Sec.~\ref{subsubsec:supercrit} leads to the conclusion that
the appearance at lower fields of {\it any\/} rotationally non-invariant
state should give non-Arrhenius switching behavior in the ring geometry.
Experimentally, what is then required is a symmetric enough ring so that
the `domain wall' part of the transition state (centered at $s_0$ in our
instanton solution) has more or less equal probability of nucleating
anyplace along the ring.  This `Goldstone mode', arising from the
rotationally invariant geometry, is ultimately where the non-Arrhenius
factor comes from.  Although the size constraints on the ring parameters
leading to the specific instanton solution~(\ref{eq:instanton}) are
difficult to realize at the present time, the generality of the basic
physical features determining the transition should lead to the predicted
crossover from Arrhenius to non-Arrhenius behavior in at least some ring
geometries that are outside of the scaling regime considered here.

\section{Acknowledgments}

This research was partially supported by NSF grants PHY-0351964 (DLS),
FRG-DMS-0101439 and DMR-0405620 (ADK), and the Akademisches Auslands Amt
and the Evangelisches Studienwerk e.V. Villigst (KM). KM thanks the Physics
Dept.~of the University of Arizona for their hospitality and support during
her stay.  DLS thanks Bob~Kohn and Valeriy Slastikov for numerous valuable
discussions, and for providing several references.

\bibliographystyle{prsty} 
\bibliography{general}

\begin{thebibliography}{10}
\bibitem{notemixing}
The magnetostatic energy also contains a mixing term between bulk and boundary
  contributions, which scales the same as the bulk term; it is therefore
  omitted here. Its contribution is the same order as that of the bulk term,
  which is shown in the text to be at least an order of magnitude smaller than
  all of the other terms.
\end{thebibliography}
%

\end{document}